# Nondestructive 3D Quantification of the Magnitude and Extent of Indentation-Induced Densification in Glass

Sang-min Lee[a], Dohyeon Lee[a,b], Juheon Lee[a,b], Herve Hugonnet[a,b], and YongKeun Park[a,b,c,*]

[a] Department of Physics, Korea Advanced Institute of Science and Technology (KAIST), Daejeon 34141, Republic of Korea

[b] KAIST Institute for Imaging, KAIST, Daejeon 34141, Republic of Korea

[c] Tomocube Inc., Daejeon 34109, Republic of Korea

[*] Corresponding author. E-mail: yk.park@kaist.ac.kr

Abstract

Vickers hardness characterizes resistance to indentation but does not distinguish the magnitude of local densification from the volume of glass affected. Here, low-coherence holotomography reconstructs three-dimensional refractive-index (RI) images of Vickers-indented fused silica, borosilicate, and soda-lime glass across applied loads and nominal dwell times. Analysis of RI changes using the Lorentz–Lorenz relation quantifies densification, while segmentation measures the spatial extent of the densified region. Cracks are observed within the same reconstructed RI images. Densified volume increased with load even when peak densification plateaued, and glasses with similar hardness showed different densification responses. Longer nominal dwell times were associated with larger volumes above fixed RI thresholds and comparatively small changes in peak densification. Joint analysis of densification and crack morphology thus characterizes subsurface indentation damage beyond hardness alone. Densification is quantified as an optical estimate under the Lorentz–Lorenz assumptions; residual stress, polarizability, and reconstruction can also affect RI.

## 1. Introduction

Glass components encounter concentrated contact loads during manufacture and use. Through-glass-via interposers and glass-core substrates, for example, undergo dicing, handling, polishing, and assembly[1, 2], and contact-induced cracks can reduce their strength[3, 4]. Indentation and abrasion therefore remain important reliability concerns for commercial glass[5]. The Vickers test provides a controlled sharp-contact experiment, reporting hardness from the applied force and residual impression size (Fig. 1a)[6]. This scalar measurement, however, does not describe the complete subsurface deformation and crack field.

Under sharp contact, glass deforms through densification, which reduces the network volume, and shear flow, which redistributes material approximately at constant volume[6, 7]. Their relative contributions depend on glass structure and correlate with Poisson's ratio, $\nu$: low-$\nu$ glasses such as fused silica generally accommodate more deformation by densification than higher-$\nu$ glasses such as soda-lime[6–8]. These deformation processes also influence radial, median, lateral, and cone or ring cracking[9–12]. Local densification and its spatial extent are distinct properties: a small region can densify strongly, whereas a larger region can densify more modestly. Measuring both is therefore necessary to distinguish subsurface responses that yield similar hardness.

Spatially resolved methods have revealed several aspects of indentation damage. Raman spectroscopy maps densification through calibrated spectral shifts, including on cross-sections[13, 14], and has been combined with constitutive finite-element analysis to characterize the field beneath fused-silica indents[15]. Atomic force microscopy measures surface pile-up and sink-in[7]; Brillouin microscopy provides depth-resolved information related to densification in silica[16]; and X-ray tomography resolves subsurface crack networks in three dimensions[17, 18]. A further method probes local densification at nanoindentation sites with nanometre-scale resolution[19]. These approaches motivate complementary measurements of the largest local change and the amount of material affected by a given indentation condition.

Quantitative phase imaging offers a nondestructive optical approach based on phase-delay measurements[20]. Diffraction phase microscopy has resolved nanoscale material topography and dynamics[21], including the axial expansion of palladium microdisks during hydrogen exposure[22], while spectrally multiplexed reflection intensity holography enables single-shot wide-field surface topography[23]. Holotomography (HT), also known as optical diffraction tomography or digital holographic tomography, extends optical characterization to volumetric refractive-index (RI) distributions reconstructed from multiple holographic or intensity measurements[24, 25]. Materials applications include the morphology and internal structure of two-photon-polymerized microstructures[26] and RI modulations in volume holographic gratings[27]. For indented glass, reconstructed three-dimensional RI images provide RI changes for densification quantification and reveal neighboring cracks within the same volume. Sung et al. established three-dimensional RI mapping of densified regions and subsurface cracks beneath Vickers indents, including load-dependent peak contrast and composition-dependent spatial profiles[28].

Here, we use low-coherence HT to reconstruct three-dimensional RI images of Vickers-indented fused silica, borosilicate, and soda-lime glass and examine how the magnitude and spatial extent of densification are related, and whether hardness captures their joint response. The Lorentz–Lorenz relation[35] quantifies densification from reconstructed RI changes. Adaptive segmentation measures the elevated-RI region in the load and composition comparisons, while fixed RI thresholds quantify its extent in the dwell-time series. Comparing these descriptors across applied load, nominal dwell time, and glass composition reveals that the affected region can expand even when local densification changes little, and that similar hardnesses can conceal distinct subsurface responses. Cracks observed in the same reconstructed RI images provide the morphological context for these quantitative comparisons. Together, peak densification, densified volume, and crack morphology provide complementary information about subsurface damage beyond hardness.

## 2. Materials and Methods

### 2.1 Glass samples

Three commercial silicate glasses were selected: borosilicate cover glass (Matsunami Glass, Japan), soda-lime slide glass (Marienfeld, Germany), and a fused-silica plate (Sungho Sigma, Republic of Korea). Table 1 lists their thicknesses, nominal refractive indices, representative Poisson's ratios, and nominal compositions. The tabulated index is the glass-specific reference $n_0$ used in Section 2.4. The Poisson's ratios provide nominal context for the composition comparison and were not measured on these specimens. Manufacturer specifications support the soda-lime value (0.23) and a representative borosilicate value (0.21; see Table 1 source notes). The fused-silica value (approximately 0.17) is consistent with the material parameters used by Bruns et al.[15].

Table 1. Glass specimens and nominal properties. The nominal reference index $n_0$ is assigned to the D line (589.3 nm) for fused silica and soda-lime and the e line (546.1 nm) for borosilicate; these values were not measured at the 444 nm illumination wavelength. Poisson's ratios are representative values, not measurements on these specimens.

| Glass | Supplier / product | Thickness (mm) | Nominal RI | Poisson ratio $\nu$ | Nominal composition (wt%) |
|---|---|---|---|---|---|
| Fused silica | Sungho Sigma / quartz square plate | 1 | 1.4580 ($n_D$) | ~0.17 | $SiO_2$ 99.9 |
| Borosilicate | Matsunami Glass / cover glass | 0.15 | 1.5250 ($n_e$) | ~0.21 | Not disclosed by supplier |
| Soda-lime | Marienfeld / slide glass | 1 | 1.5200 ($n_D$) | 0.23 | $SiO_2$ 72.6, $Na_2O$ 13.0, CaO 8.8, MgO 4.3, $Al_2O_3$ 0.6, $K_2O$ 0.3, $SO_3$ 0.2, $Fe_2O_3$ 0.1 |

Sources and notes: Marienfeld, Microscope Slides datasheet (September 2020), reports the soda-lime oxide composition, $n_D$ = 1.52, and Poisson's ratio 0.23. The Matsunami Fine Glass catalogue (8th edition, 2016) lists Poisson's ratio 0.21 for #0100 and D263LA glasses; it is used as representative context for the borosilicate specimen, whose composition was not disclosed. The Matsunami RCM cover-glass specification reports $n_e = 1.5255 \pm 0.0015$, consistent with the adopted 1.5250. For fused silica, the RI is the nominal analysis reference and the Poisson's ratio follows the literature cited in Section 2.1.

## 2.2 Vickers indentation

Indents were produced with a microhardness tester (HVS-1000, SINOWON, China) equipped with a Vickers diamond indenter (Fig. 1a). Two series were acquired for each glass. The load series used eight peak forces (0.098, 0.245, 0.490, 0.981, 1.962, 2.942, 4.904, and 9.808 N) at a nominal 15 s dwell. The dwell series used seven nominal dwell times (0–60 s in 10 s increments), at 9.808 N for borosilicate and soda-lime and 1.962 N for fused silica. The lower fused-silica load limited excessive cracking. Indentation was performed at room temperature.

Ten replicate indentation experiments were performed for each condition in each glass. Indents too severely damaged to measure were excluded from analysis. Vickers hardness was calculated from the applied force $P$ and the mean diagonal length $d$ of the residual impression, measured from the HT reconstruction. With $P$ in newtons and $d$ in micrometers, the hardness in gigapascals is $H_V = 1854.4\, P/d^2$.

## 2.3 Three-dimensional RI reconstruction by low-coherence HT

Three-dimensional RI images were reconstructed using a low-coherence HT system (HT-T1D, Tomocube Inc., Republic of Korea; Fig. 1b). Low-coherence illumination is useful for resolving weak RI variations because coherence-related speckle affects digital holography and optical coherence tomography[29, 30]. Partially coherent and broadband approaches extend diffraction tomography beyond conventional coherent illumination[31–34]. Light-emitting diode (LED) illumination was programmed with a digital micromirror device (DMD). Intensity stacks acquired under four illumination patterns at each axial position were processed by phase deconvolution[36–38] to reconstruct the three-dimensional RI distribution (Fig. 1c,d). These reconstructed images were then analyzed to quantify densification from RI changes and observe cracks. The optical configuration comprised a 40× air objective (NA 0.95), a condenser with NA 0.72, and an LED centered at 444 nm (approximately 20 nm full width at half maximum). The axial acquisition step was 1 µm, and the nominal lateral field of view was 218 × 165 µm.

Each specimen was immersed in index-matching oil with a nominal RI of 1.4450 for fused silica and 1.5100 for both borosilicate and soda-lime glass. This reduced the glass–medium contrast while retaining contrast in the deformed region and at cracks. Reconstruction settings, including total-variation regularization, were held fixed across glasses, loads, and nominal dwell times. The reconstruction used a weak-scattering approximation.

## 2.4 Densification quantification from reconstructed RI changes

The RI of vitreous silica changes under pressure, including in a regime without retained densification[39]. At higher pressures, RI measurements probe coupled changes in density, structure, and polarizability[35]. The Lorentz–Lorenz relation connects refractive index $n$ with mass density $\rho$ through the specific refraction $K$,

$$\frac{n^2-1}{n^2+2} = K\,\rho, \tag{1}$$

where $K$ is the specific refraction, with units of m$^3$ kg$^{-1}$ when mass density is expressed in kg m$^{-3}$. Dividing by the corresponding relation for the undeformed reference gives

$$\frac{\rho}{\rho_0} = \frac{(n^2-1)(n_0^2+2)}{(n^2+2)(n_0^2-1)}, \tag{2}$$

Here, $n_0$ and $\rho_0$ describe the undeformed reference state. Equation (2) assumes unchanged specific refraction $K$ and a common wavelength and temperature for $n$ and $n_0$. Structural changes and residual stress can also alter RI. Moreover, the nominal D/e-line indices in Table 1 are not independently calibrated references at 444 nm; fused-silica dispersion is well established[40]. We quantify densification from RI changes as a Lorentz–Lorenz optical estimate, $\Delta\rho/\rho_0 = \rho/\rho_0 - 1$, rather than an independently calibrated density.

2.5 RI image analysis for densified volume and crack observation

Each reconstructed three-dimensional RI image was analyzed to segment the elevated-RI region for densified-volume quantification and to observe cracks (Fig. 1e). Triangle thresholding of the smoothed RI histogram[41, 42] provided an initial mask. Disconnected high-RI components were removed, and an object center was estimated in each slice. RI gradients, local contrast, and radial displacement identified boundary candidates along polar radial profiles. A smooth closed contour was selected in each slice; these contours were stacked into a three-dimensional mask and trimmed axially to remove reconstruction artifacts. Segmentation parameters were held fixed across conditions. Cracks were observed as sub-baseline RI regions in the reconstructed images and identified by their morphology.

The adaptive mask follows the RI distribution rather than a fixed RI isosurface. For the dwell series, we additionally measured volumes above two fixed RI thresholds for each glass: 1.5450 and 1.5550 for borosilicate, 1.5250 and 1.5350 for soda-lime, and 1.4800 and 1.4900 for fused silica. Each cutoff remained fixed within its series. The threshold pairs do not represent equivalent density cutoffs across glasses. Low-RI crack voxels do not contribute to these high-RI volumes.

Four quantitative descriptors of densification and its spatial extent were extracted from the reconstructed RI images. Local densification is denoted by $\varepsilon = \Delta\rho/\rho0$. Peak densification, $\varepsilon_{\mathrm{max}}$, and mean densification, $\overline{\varepsilon}$, denote the maximum and spatial mean of local densification within the adaptive mask, respectively. Densified volume denotes the volume of the adaptive segmentation mask, not the volume lost by compaction. These three descriptors were used for the load and composition comparisons. Fixed-threshold volume denotes the volume above a specified RI cutoff and was used for the dwell comparison. Fractional volume change was calculated as $[V(t) - V(0)]/V(0)$, where t is nominal dwell time and V is the arithmetic mean volume at that condition. This is a ratio of condition means, not a mean of paired within-indent changes.

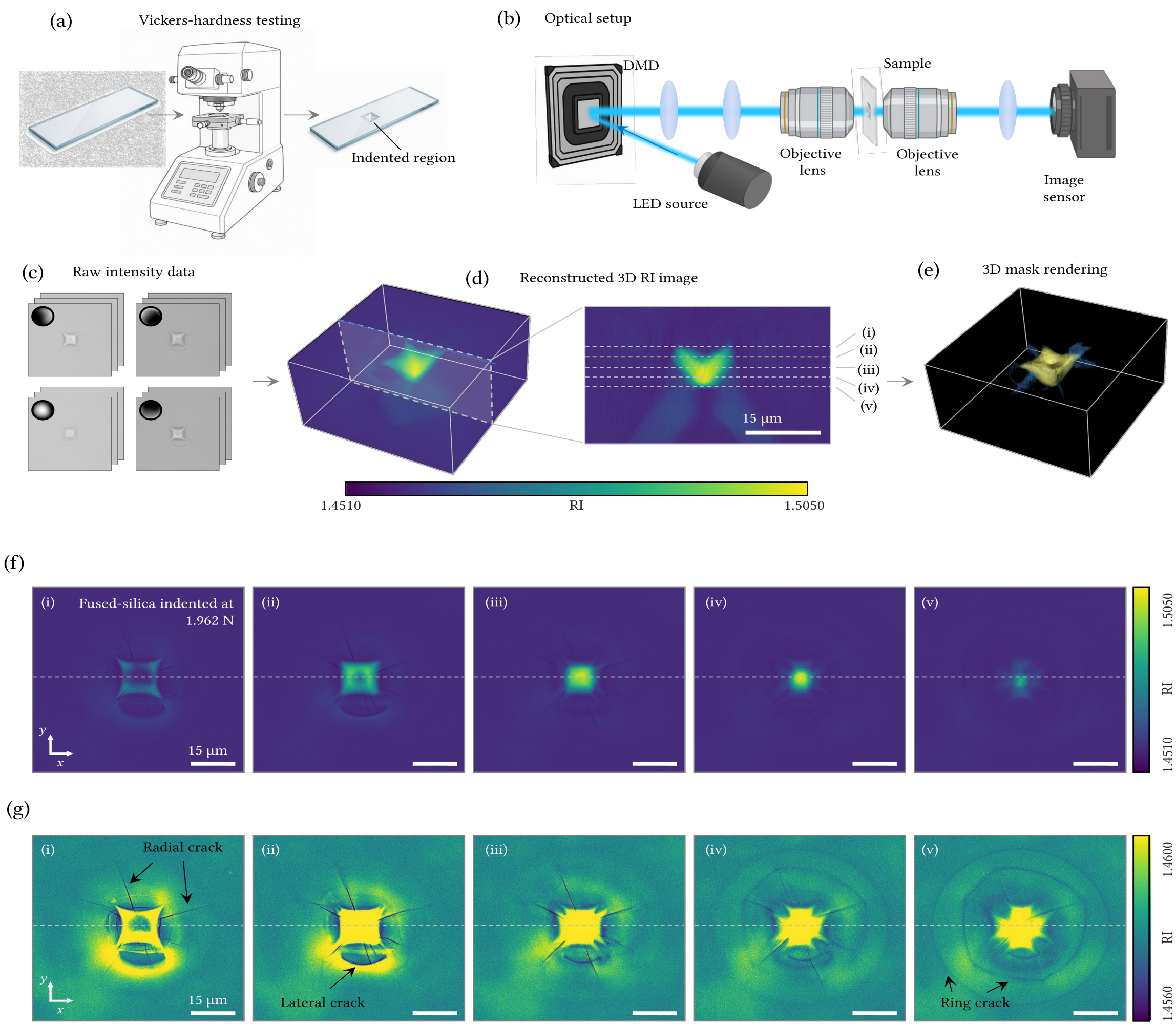

Figure 1. Three-dimensional RI reconstruction and image analysis for densification quantification and crack observation in Vickers-indented glass. (a) Vickers indentation. (b) Optical setup with LED illumination, a digital micromirror device (DMD), an index-matched specimen, objective lenses, and an image sensor. (c) Intensity images acquired under different illumination patterns. (d) Reconstructed RI volume and *xz* section marking slice positions (i)–(v). (e) Three-dimensional rendering of the segmented densified region and cracks observed in the RI image. (f) *xy* sections through a representative fused-silica indent at 1.962 N, displayed over RI 1.4510–1.5050. (g) Corresponding sections displayed over RI 1.4560–1.4600 to emphasize radial, lateral, and ring crack channels. Scale bars, 15 µm.

## 2.6 Statistics

Measurements are summarized as arithmetic means with standard errors of the mean (SEM). Ten replicate indentation experiments were performed per condition, with severely damaged indents excluded as described in Section 2.2. Comparisons are descriptive. Each composition was represented by one specimen; therefore, variation among indents does not quantify variation among independent specimens. Fractional volume changes were calculated from condition means as defined in Section 2.5.

## 3. Results and Discussion

### 3.1 Three-dimensional imaging of densified regions and cracks

Figure 2 shows representative three-dimensional RI reconstructions of Vickers-indented glass after unloading. The segmented high-RI regions, rendered in yellow, delineate the densified cores, while the blue low-RI regions reveal cracks identified by their morphology. The fused-silica series in Fig. 2a spans forces from 0.098 to 9.808 N at a nominal 15 s dwell, showing how the spatial arrangement of the densified region and surrounding cracks changes with load.

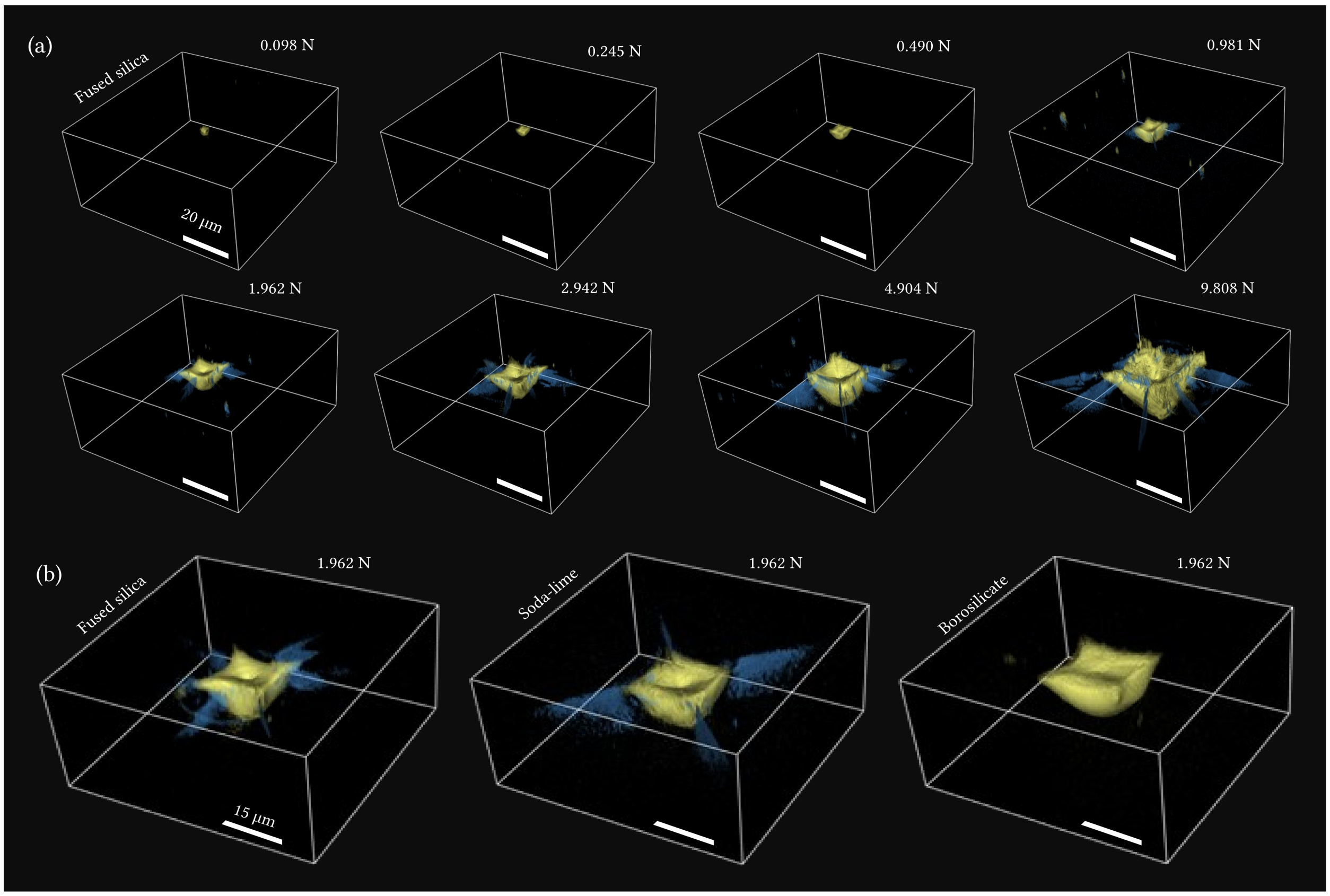

Figure 2. Reconstructed three-dimensional refractive-index (RI) images showing densified regions and cracks across load and glass composition. (a) Fused-silica indents produced at a nominal 15 s dwell and forces of 0.098, 0.245, 0.490, 0.981, 1.962, 2.942, 4.904, and 9.808 N. Scale bar, 20 µm. (b) Fused silica, soda-lime, and borosilicate at a matched force of 1.962 N. Scale bar, 15 µm. Yellow denotes the segmented high-RI region; blue denotes low-RI regions identified morphologically as open cracks in the reconstructed images.

At the lowest forces, 0.098–0.490 N, the fused-silica reconstructions show a small, localized high-RI core. The core expands as the applied force increases, and distinct low-RI crack channels become apparent at approximately 1 N. At higher forces, these channels extend outward from the enlarged core and become more prominent in the representative reconstructions. The images therefore resolve both the growth of the densified region and its spatial relationship to neighboring cracks.

The matched-force comparison at 1.962 N reveals composition-dependent differences in core geometry and crack morphology (Fig. 2b). Fused silica shows a lobed high-RI core with crack channels extending in several directions. Soda-lime also shows pronounced low-RI features extending laterally from the core, whereas borosilicate shows a smoother, more continuous high-RI region without similarly prominent crack channels in this representative image. These differences establish the three-dimensional morphological context for comparing the magnitude and spatial extent of densification.

3.2 Load and composition dependence of densification and densified volume

Analysis of the reconstructed RI images quantifies how the magnitude and spatial extent of densification vary with applied load (Fig. 3a–c). Peak densification, $\varepsilon_{\max}$, and mean densification, $\overline{\varepsilon}$, are obtained from RI changes, and densified volume is obtained by segmenting the elevated-RI region. Peak densification rises steeply below approximately 1 N and more slowly thereafter (Fig. 3a). It continues to increase up to 9.808 N in fused silica and borosilicate, whereas soda-lime reaches a plateau near 1 N. Mean densification follows a similar pattern, except that the soda-lime mean peaks near 1–2 N and decreases slightly at higher loads (Fig. 3b). The densified volume increases by approximately 2.5 orders of magnitude across the load range in all three glasses (Fig. 3c). Soda-lime has a larger volume than borosilicate below approximately 1 N, but borosilicate has the larger volume at higher loads; fused silica has the smallest volume throughout. In soda-lime, the peak plateau and declining mean accompany continued growth of the elevated-RI region. The load response therefore changes both the extent of this region and its average densification, which a peak measurement alone does not capture.

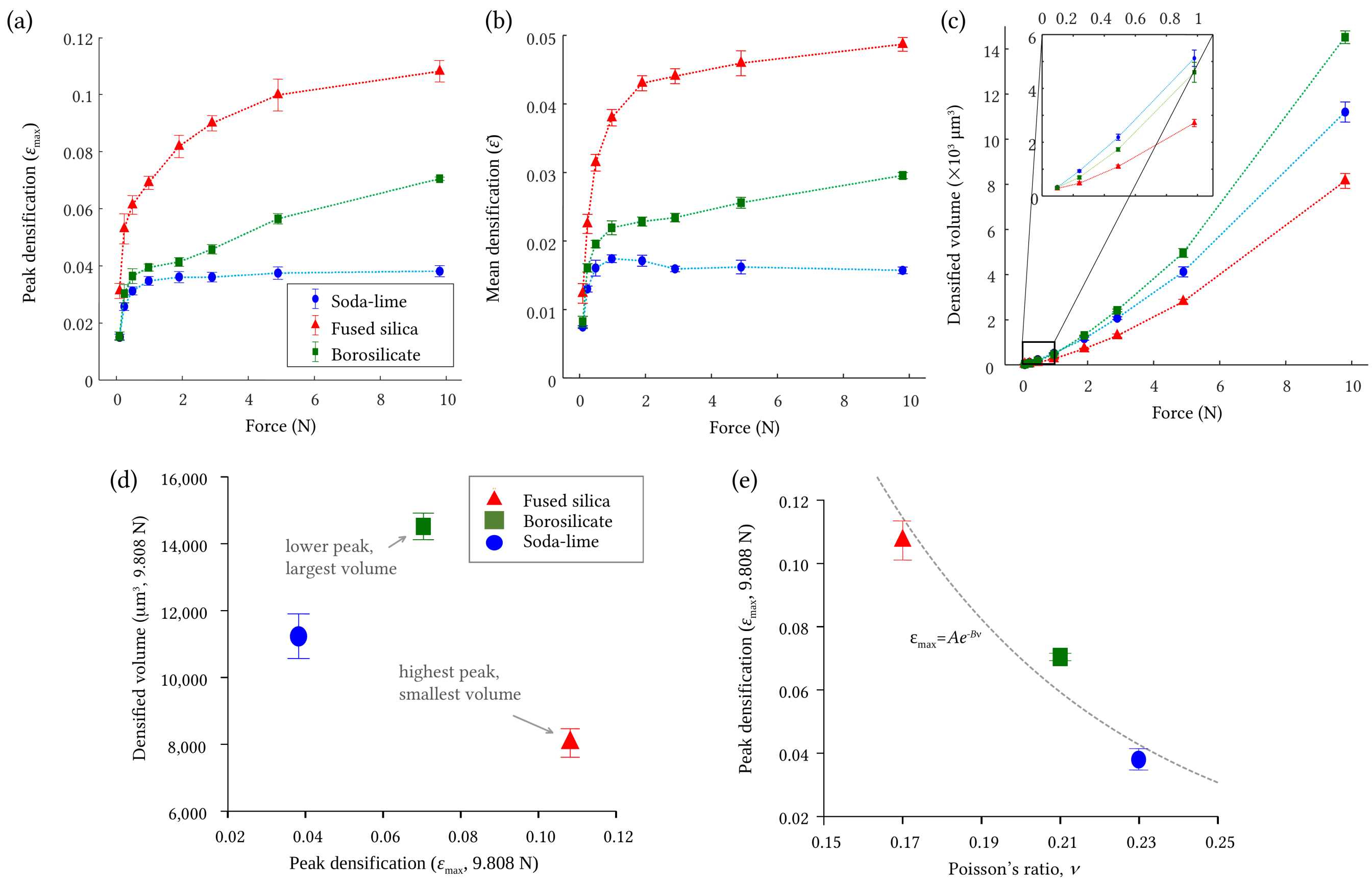

Figure 3. Load and composition dependence of densification and densified volume. (a) Peak densification, $\varepsilon_{\max}$, and (b) mean densification, $\overline{\varepsilon}$, versus force. These dimensionless quantities are the maximum and spatial mean of Δρ/ρ0 within the adaptive mask, respectively, quantified from reconstructed RI changes using the Lorentz–Lorenz relation. (c) Densified volume, defined as the volume of the adaptive segmentation mask, versus force; the inset expands 0–1 N. (d) Densified volume versus peak densification at 9.808 N. (e) Peak densification versus representative Poisson's ratio at 9.808 N. The dashed exponential guide and its digitized coefficients are discussed in Section 3.2. Error bars indicate SEM.

Sung et al. reported a high-load decrease in the soda-lime peak associated with a subsurface median crack crossing the high-RI region[28]. Here, the peak plateaus while the densified volume continues to grow. Both observations show why a peak value alone cannot describe the spatial response to load. The present analysis quantifies the volume alongside the peak and extends the comparison to nominal dwell time. Differences in specimens, cracking, and measurement conditions prevent attribution of the different peak trends to illumination coherence alone.

Growth of the affected region does not require a proportional increase in its peak contrast. Constitutive descriptions of silica couple compaction, shear deformation, and densification hardening[43], allowing additional material to enter the deformation zone as load increases. The soda-lime response is consistent with this distinction, although an optical plateau alone does not establish a saturation density. Likewise, the annealing-recoverable densification contribution can decrease with load in some glasses[44] while the affected region expands. Segmented volume measures the spatial extent selected by the adaptive mask; determining the volume lost by compaction would instead require a calibrated density or volumetric-strain field.

Borosilicate and soda-lime had similar Vickers hardnesses at 9.808 N: 5.096 ± 0.011 and 5.230 ± 0.009 GPa, respectively (mean ± SEM; 10 diagonal pairs per glass). Nevertheless, their peak densifications and densified volumes differed. Hardness and optical descriptors are compared at the specimen level rather than through a paired analysis of individual indents. The distinction between hardness and the contributions of densification and plastic flow also has precedent in studies of silicate glasses[45].

Figure 3d compares densified volume with peak densification across the three glasses at 9.808 N. At this load, fused silica has the highest peak densification but the smallest densified volume; borosilicate has an intermediate peak and the largest volume; and soda-lime has the lowest peak and an intermediate volume. The peak ordering is inverse to the ordering of the representative Poisson's ratios (Fig. 3e). The volume ordering is different: the glass with the highest RI-derived peak densification does not have the largest densified volume.

The decrease in peak densification with increasing $\nu$ is consistent with the connection between glass-network packing and densification capacity. Low-$\nu$ glasses such as fused silica have relatively open networks with greater scope for compaction. Rouxel et al. correlated Poisson's ratio with the maximum density increase retained after hydrostatic compression and decompression[8]. This connection helps interpret the local peak, but it does not by itself rank the spatial extent of deformation under a sharp contact. Glass composition and loading history also influence the balance between densification and shear flow[6].

The dashed guide in Fig. 3e follows $\varepsilon_{\max} = A\exp(-B\nu)$, where $\varepsilon_{\max}$ is the dimensionless peak densification quantified from RI changes as defined in Section 2.4. Values of $A \approx 1.80$ and $B \approx 16.2$ were estimated by fitting points digitized from the displayed guide, rather than by fitting individual indentation measurements. $A$ is the extrapolated intercept at $\nu = 0$, outside the measured range, and $B$ describes the magnitude of the logarithmic slope. The curve summarizes the ordering of three compositions; it does not establish a constitutive law or universal coefficients.

Rouxel et al. reported an exponential relation for the density increase retained after hydrostatic compression and decompression, with $A \approx 1.50$ and $B \approx 13$ in the same fractional convention[8]. The coefficients are comparable in scale, but the loading conditions and measured quantities differ. As an independent indentation benchmark, Kassir-Bodon et al. reported a peak densification of approximately 3.8% in soda-lime-silicate window glass under 10 and 20 N Vickers loading[14]. These precedents support the order of magnitude of the optical estimate, without calibrating the present specimens.

Poisson's ratio is also an incomplete description of a commercial glass. Malchow et al. showed that composition and cooling history affect plastic deformation, densification, and cracking in sodium borosilicate glasses[46]. The nominal ratio used here therefore provides comparative context, not a specimen-specific predictor. Additional compositions and thermal states would be needed to generalize the observed ordering.

### 3.3 Dwell-time dependence of peak densification and threshold volume

For each glass, three-dimensional RI images of indents were analyzed at a fixed force over nominal dwell times of 0–60 s (Fig. 4). Borosilicate and soda-lime were tested at 9.808 N and fused silica at 1.962 N to limit excessive cracking. Each time point represents different indents imaged after unloading, rather than continuous observation of a single indent.

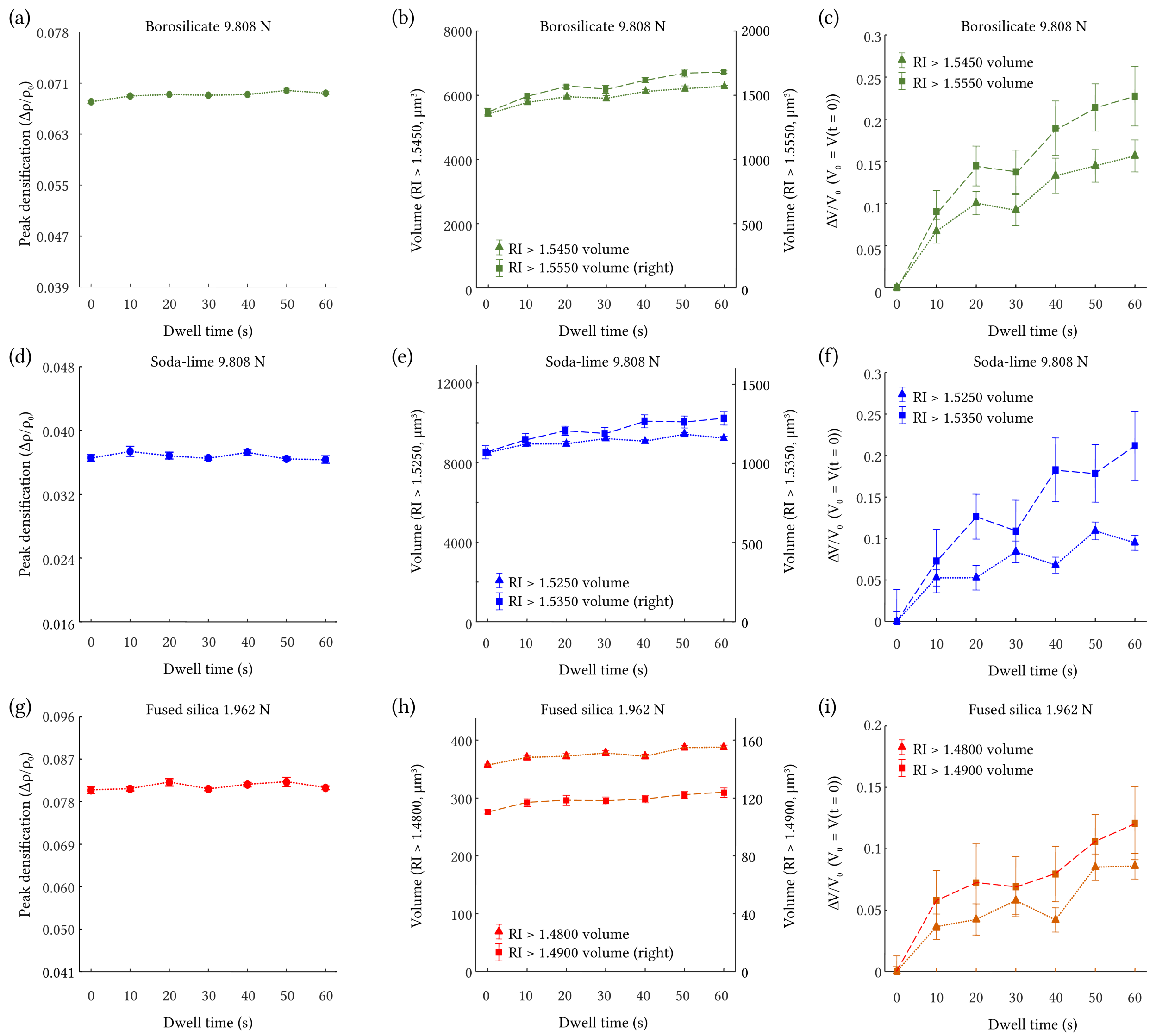

Figure 4. Peak densification and fixed-RI threshold volume versus nominal dwell time. Rows show borosilicate at 9.808 N (a–c), soda-lime at 9.808 N (d–f), and fused silica at 1.962 N (g–i). (a,d,g) Peak densification, quantified from reconstructed RI changes and expressed as $\Delta\rho/\rho_0$. (b,e,h) Volumes above fixed RI thresholds of 1.5450 and 1.5550 for borosilicate, 1.5250 and 1.5350 for soda-lime, and 1.4800 and 1.4900 for fused silica. Lower- and higher-threshold volumes use the left and right axes, respectively. (c,f,i) Fractional volume changes, $[V(t) - V(0)]/V(0)$, calculated from condition means. Each nominal dwell time represents different indents measured after unloading. Error bars indicate SEM.

Peak densification varied only weakly with nominal dwell time (Fig. 4a,d,g). Fused silica and soda-lime showed little systematic change, whereas borosilicate showed a small increase. The borosilicate peak at 60 s was approximately 2% higher than at 0 s. This is a relative change in the densification estimate, not a percentage-point change in density.

Fixed-threshold volumes showed a net increase between 0 and 60 s in all three glasses, although the trajectories were not monotonic (Fig. 4b,c,e,f,h,i). The lower- and higher-threshold mean volumes increased by 15.8% and 22.5% in borosilicate, 9.5% and 21.2% in soda-lime, and 8.6% and 12.1% in fused silica. The higher-threshold endpoint increases therefore ranged from approximately 12% to 23%.

Peak densification and fixed-threshold volume are obtained by analyzing the same reconstructed RI images: the former quantifies the largest RI-derived densification, and the latter measures how much material exceeds a selected RI. The larger endpoint response at the higher threshold indicates a change in the distribution of recovered RI values. It does not by itself distinguish physical redistribution from RI shifts near a cutoff or reconstruction effects.

Lai et al. reported lower nanohardness and larger densified and shear-flow volumes with longer dwell times in silica and soda-lime silicate glass, using nanoindentation, AFM, and annealing[47]. Their densification volume measures annealing-recoverable deformation; the present threshold volume measures the extent of elevated RI. Duan et al. found that slower loading enhanced subsurface densification in soda-lime silicate glass while shear flow changed comparatively little[48]. These results show why both glass composition and loading history matter, while dwell time and loading rate must be treated as different variables.

Continued penetration during a dwell can increase the contact size, and deformation or recovery can modify the RI distribution before imaging. The present analysis quantifies densification and elevated-RI volume after unloading and relates them to nominal dwell time, but does not separate these processes. Instrumented load–displacement records and controlled imaging delays would be needed to determine their individual contributions. Recovery after unloading can depend on composition and environment, as demonstrated in chemically distinct oxide glasses[49]. That observation does not establish recovery in the present specimens; it emphasizes the distinction between programmed dwell time and elapsed time after unloading.

The crack-free borosilicate dwell series shows that a larger threshold volume can occur without detectable cracks. Most fused-silica and soda-lime indents contained cracks and showed the same direction of volume response. Excluding low-RI crack voxels prevents their direct inclusion in the high-RI volume, but not their possible effects on the surrounding RI distribution or reconstruction. Because fused silica was tested at a lower force, cross-composition comparisons are limited to the direction of the dwell response.

### 3.4 Crack observation in reconstructed three-dimensional RI images

Cracks are observed in the reconstructed three-dimensional RI images, which show their morphology and location relative to the densified region. For the representative fused-silica indent at 1.962 N, Fig. 1d shows the three-dimensional RI volume and an *xz* section; Fig. 1f,g shows *xy* sections with two display ranges to reveal the densified core and crack channels. At 9.808 N, representative indents exhibit radial and subsurface median cracks (Fig. 5a–c). Fused silica shows a lateral crack, whereas borosilicate and soda-lime show band-like structures morphologically consistent with reported shear faults beneath pyramidal indents[50]. These assignments are morphological; crack-network statistics were not measured across replicates.

Densification and shear flow both affect the stresses that drive cracking. Kato et al. associated greater densification with higher resistance to Vickers crack initiation and attributed this to reduced residual stress[51]. Constitutive calculations by Barthel et al. instead emphasized shear flow stress as a determinant of the residual stress field and shear localization as a possible route to crack initiation[52]. The reconstructed RI images show the spatial relationship between observed cracks and densified regions, but they do not identify a causal sequence or directly rank fracture resistance.

Crack initiation and crack propagation are also distinct. In sodium aluminoborosilicate glasses, Januchta et al. found no simple correlation between resistance to indentation crack initiation and fracture toughness[53]. Peak densification and densified volume may help characterize the associated deformation, but neither is itself a measure of fracture toughness.

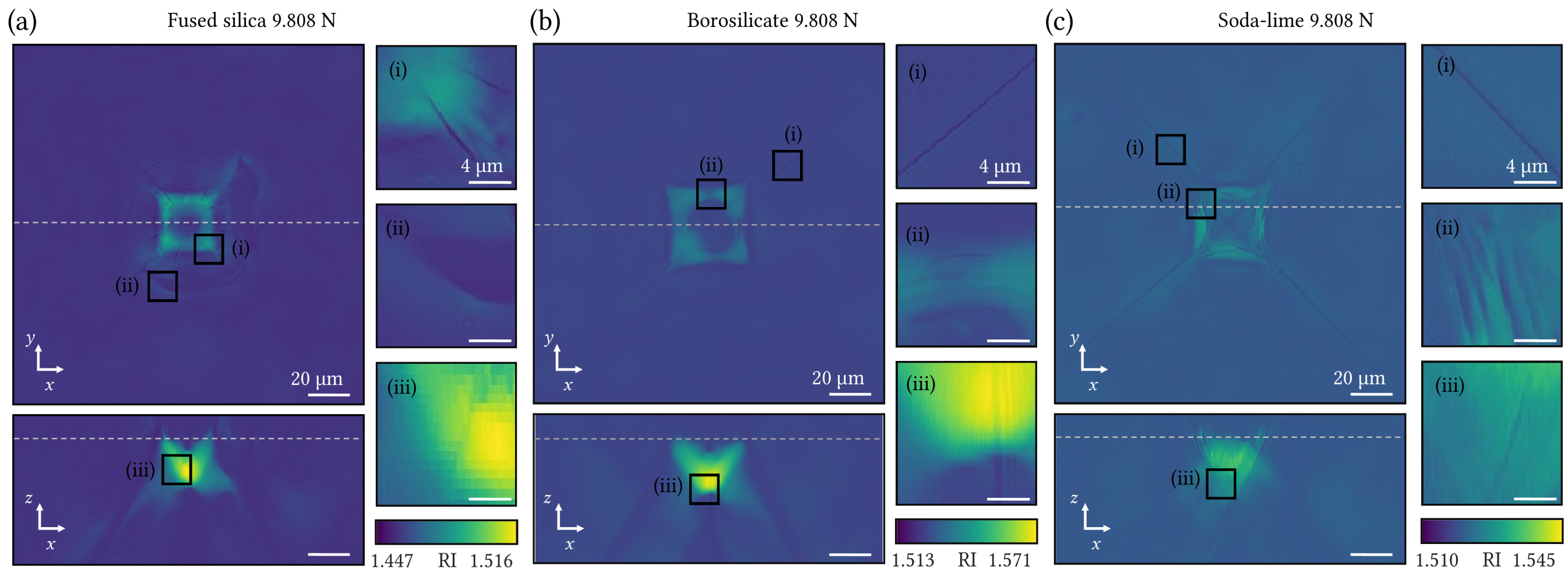

Figure 5. Cross-sections of reconstructed three-dimensional refractive-index (RI) images showing cracks in representative indents at 9.808 N. (a) Fused silica, (b) borosilicate, and (c) soda-lime, shown in *xy* (top) and *xz* (bottom). Display ranges are 1.4470–1.5160, 1.5130–1.5710, and 1.5100–1.5450, respectively. Insets enlarge (i) radial cracks in *xy*, (ii) a lateral crack in fused silica and band-like structures in borosilicate and soda-lime, and (iii) subsurface median cracks in *xz*. Feature assignments are based on morphology. Scale bars, 20 μm; inset scale bars, 4 μm.

## 3.5 Limitations and outlook

Densification quantification from RI changes and crack observation in the reconstructed images have two main interpretation limits. First, converting RI into density requires a common reference scale and unchanged specific refraction, while residual stress can also contribute to RI. Second, limited angular coverage and regularization can bias both the recovered contrast and shape[54, 55]. Independent RI or Raman calibration and reconstruction-sensitivity tests would strengthen the mechanical interpretation; using the same reconstruction settings alone does not ensure equal bias across specimens. The dwell series was not load-matched across all glasses, and post-unloading images do not resolve deformation and recovery in time.

Measurements during loading can provide complementary structural information. Using in-situ small-angle X-ray scattering during wedge indentation of vitreous silica, Fuhrmann et al. distinguished reversible changes in electron-density fluctuations from persistent changes near the indentation center[56]. Although these observables differ from RI contrast, the study illustrates why RI images acquired after unloading capture only part of the structural response to indentation.

Polarization-resolved tomography provides a complementary route to optical anisotropy. Dielectric tensor tomography reconstructs principal refractive indices and optical-axis orientations[57] and has characterized polarization holographic elements[58]; incoherent dielectric tensor tomography extends this capability to biaxial anisotropy[59]. These methods could help constrain anisotropic optical contributions, although anisotropy alone would not separate isotropic density changes from hydrostatic stress.

## 4. Conclusions

Low-coherence HT reconstructed three-dimensional RI images beneath Vickers indents. Analysis of RI changes using the Lorentz–Lorenz relation quantified densification, segmentation measured the densified volume, and cracks were observed in the same reconstructed images. Borosilicate and soda-lime exhibited different peak densifications and densified volumes despite similar hardness. Densified volume continued to grow with load after the soda-lime peak densification plateaued, and longer nominal dwell times were associated with larger fixed-threshold volumes with comparatively small changes in peak densification. Quantifying densification and observing crack morphology within the same three-dimensional RI images therefore provides a more complete description of subsurface indentation damage.

The RI-derived densification estimates remain subject to the Lorentz–Lorenz assumptions and RI-reference limitations, and crack observation does not directly measure fracture resistance. Within this scope, analysis of reconstructed three-dimensional RI images extends conventional hardness testing by quantifying the magnitude and spatial extent of densification and revealing associated cracks. Independent density calibration, surface-deformation measurements, and crack-resistance tests would further establish the value of these image-derived measurements for assessing contact damage in glass.

Acknowledgements

This work was supported by the Samsung Research Funding Center of Samsung Electronics under Grant SRFC-IT2501-05 and by National Research Foundation of Korea (NRF) grants funded by the Korea government (MSIT) (RS-2024-00442348, RS-2022-NR068141).

CRediT authorship contribution statement

Sang-min Lee: Methodology, Software, Validation, Formal analysis, Investigation, Data curation, Visualization, Writing – original draft. Dohyeon Lee: Software, Investigation. Juheon Lee: Investigation, Formal analysis. Herve Hugonnet: Methodology, Software, Supervision, Writing – review & editing. YongKeun Park: Conceptualization, Methodology, Resources, Supervision, Project administration, Funding acquisition, Writing – review & editing.

Declaration of competing interest

YongKeun Park is a co-founder and Chief Executive Officer of Tomocube Inc. (Daejeon, Republic of Korea), the manufacturer of the holotomography system used in this study, and holds equity in the company. Sang-min Lee, Dohyeon Lee, Juheon Lee, and Herve Hugonnet declare no competing financial interests.

Data availability statement

The data that support the findings of this study are available from the corresponding author upon reasonable request.